\documentclass[twocolumn,english,pre]{revtex4}
\usepackage[T1]{fontenc}
\usepackage[latin1]{inputenc}
\usepackage{graphicx}
\usepackage{amssymb}

\makeatletter

\usepackage{graphicx}
\usepackage{graphicx}

\usepackage{epsf}

\usepackage{babel}
\makeatother
\begin{document}

\title{Aging Correlation Functions for Blinking Nano-Crystals, and Other
On - Off Stochastic Processes }

\author{G. Margolin and E. Barkai}

\affiliation{Department of Chemistry and Biochemistry, Notre Dame University,
Notre Dame, IN 46556.}

\date{\today{}}

\begin{abstract}
Following recent experiments on power law blinking behavior of single
nano-crystals, we calculate two-time intensity correlation functions
$\langle I(t)I(t+t')\rangle$ for these systems. We use a simple two
state (\emph{on} and \emph{off}) stochastic model to describe the
dynamics. We classify possible behaviors of the correlation function
and show that aging, e.g., dependence of the correlation function
on age of process $t$, is obtained for classes of the \emph{on} time
and \emph{off} time distributions relevant to experimental situation.
Analytical asymptotic scaling behaviors of the intensity correlation
in the double time $t$ and $t'$ domain are obtained. In the scaling
limit $\langle I(t)I(t+t')\rangle\rightarrow h(x)$, where four classes
of behaviors are found: (i) finite averaged \emph{on} and \emph{off}
times $x=t'$ (standard behavior) (ii) \emph{on} and \emph{off} times
with identical power law behaviors $x=t/t'$ (case relevant for capped
nano-crystals). (iii) exponential \emph{on} times and power law \emph{off}
times $x=tt'$ (case relevant for uncapped nano-crystals). (iv) For
defected \emph{off} time distribution we also find $x=t+t'$. Origin
of aging behavior is explained based on simple diffusion model. We
argue that the diffusion controlled reaction $A+B\rightleftharpoons AB$,
when followed on a single particle level exhibits aging behavior. 
\end{abstract}
\maketitle

\section{Introduction}

The fluorescence emission of single colloidal nano-crystals (NC),
e.g. CdSe quantum dots, exhibits interesting intermittency behavior
\cite{Nirmal}. Under laser illumination, single NCs blink: at random
times the NC will turn from state \emph{on} in which many photons
are emitted, to state \emph{off} in which the NC is turned off. One
method to characterize blinking quantum dots is based on the distribution
of \emph{on} and \emph{off} times. According to the theory of Efros
and Rosen \cite{EfrosRosen97}, these \emph{on} and \emph{off} times,
correspond to a neutral and ionized NC respectively. Thus statistics
of \emph{on} and \emph{off} times teaches us on ionization events
on the level of a single NC. Surprisingly, \cite{Kuno,Ken} distributions
of \emph{on} and \emph{off} times exhibit power law statistics. For
capped NCs the probability density function (PDF) of \emph{on} time
decays like $\psi_{+}(t)\sim t^{-1-\alpha_{+}}$, while for \emph{off}
times $\psi_{-}(t)\sim t^{-1-\alpha_{-}}$, where in many cases $\alpha_{+}$
and $\alpha_{-}$ are close to $0.5$ \cite{Brokmann}. 

Statistical behavior of single emitting NCs, and more generally single
molecules \cite{MO} or atoms \cite{Wolf,Short,Plenio}, is usually
characterized based on intensity correlation functions \cite{Dahan,Verberk,Oijen}.
The calculation of intensity correlation functions, and the related
Mandel $Q$ parameter, for single molecule spectroscopy is a subject
of intense theoretical research \cite{Wang1,Geva3,Schenter,Berez,PRL,ADV,ShilongE,Vlad,Barsegov2,BrownG}
(see \cite{Annual} for review). Experiments on single NCs show how
the correlation function method yields dynamical information over
time scale from nano-second to tens of seconds \cite{Dahan}. The
correlation function of single NCs exhibits a non-ergodic behavior,
as such these systems exhibit behavior very different than other single
emitting objects. 

The goal of this paper is to calculate the averaged intensity correlation
function for the emitting NCs. For this aim we use a simple two state
stochastic model. The motivation for the calculation is twofold. First,
the averaged correlation function exhibits interesting aging behavior,
as we will demonstrate. This aging behavior is a signal of non-ergodicity.
Secondly, to obtain understanding of non-ergodic properties of the
correlation function, one must first understand how the averaged correlation
function behaves. In a future publication we will discuss the non-ergodic
behavior of the NC correlation function, namely the question of the
distribution of correlation functions obtained from single trajectory
measurements. 

Aging in our context means that the (non-normalized) intensity correlation
function \begin{equation}
C(t,t')=\langle I(t)I(t+t')\rangle,\label{eq0DEF}\end{equation}
 and the normalized correlation \begin{equation}
g^{(2)}(t,t')\equiv\frac{\langle I(t)I(t+t')\rangle}{\left\langle I(t)\right\rangle \left\langle I(t+t')\right\rangle }=\frac{C(t,t')}{\left\langle I(t)\right\rangle \left\langle I(t+t')\right\rangle }\label{eq:g2def}\end{equation}
 depend on the the age of the process $t$ even in the limit of long
times. Here $I(t)$ is the fluctuating stream of photons emitted from
the NC (units counts per second). In the ergodic phase (i.e. when
both the mean \emph{on} and \emph{off} times are finite) stationarity
is reached meaning that $C(t,t')\rightarrow C(t')$ when $t\rightarrow\infty$
and similarly for $g^{(2)}(t,t')$. The average in (\ref{eq0DEF})
is over many single NC intensity trajectories. 

Previously, Jung, Barkai and Silbey \cite{Jung} showed the relation
of the problem to the L\'{e}vy walk model \cite{Schlesinger}. The
approach in \cite{Jung} is based on the calculation of Mandel's $Q$
parameter and does not consider the aging properties of the NCs. Verberk
and Orrit \cite{Verberk} considered the problem of the correlation
function for blinking NC, however they assume that the mean \emph{on}
and the mean \emph{off} times are finite, while the experiments show
an infinite \emph{off} and \emph{on} times (for capped NCs). To overcome
this problem Verberk and Orrit introduce cutoffs on the \emph{on}
and \emph{off} times. The results of Verberk and Orrit are different
than ours: they do not exhibit aging and they are meant to describe
the correlation function of a single trajectory (however the ergodic
problem was not considered). Brokmann et al. \cite{Brokmann} have
measured aging behavior of a number of NCs. They concentrate on the
measurement of the persistence probability (see details below) while
this work is devoted to the investigation of the intensity correlation
function. 

We note that concepts of statistical aging and persistence, used in
this manuscript, were introduced previously in the context of the
trap model and glassy dynamics by Bouchaud and co-workers \cite{Bouchaud,Dean,Month,Bertin}.
Statistical aging is found in continuous time random walks \cite{Cheng,Allegrini},
and in deterministic dynamics of low dimensional chaotic systems \cite{Chaos}.
Aging in complex dynamical systems, for example super-cooled liquids
or glasses is a topic of much research \cite{Leticia}. In contrast
we will later show that aging in NCs may be a result of very simple
physical processes (e.g. normal diffusion). Thus we expect aging and
non-ergodic behavior to be important in other single molecule systems. 

In the context of fractal renewal theory, Godr\`{e}che and Luck \cite{GL}
have considered the problem of the averaged correlation, however,
in the language of single NC spectroscopy, they assume that statistical
properties of the \emph{on} time are identical to the statistical
properties of \emph{off} times, i.e. $\psi_{+}(t)=\psi_{-}(t)$. Here
we use methods developed in \cite{GL} to the case relevant to experiments
$\psi_{+}(t)\ne\psi_{-}(t)$. We also obtain the aging correlation
function in the scaling limit in the time domain. 

This paper is organized as follows. In Sec. \ref{SecModel} the mathematical
model is presented and the physical meaning of \emph{on} and \emph{off}
times distributions is discussed. A brief discussion of ensemble average
and time average correlation function is given. In Sec. \ref{SecNumber}
statistical properties of the stochastic process are considered, e.g.
average number of jumps etc. In Sec. \ref{SecFor} the distribution
of the forward recurrence time is calculated, the latter is important
for the calculation of the aging correlation function. In Sec. \ref{SecRen}
we calculate probability of number of transitions between $t$ and
$t+t'$, with which the mean intensity (Sec. \ref{SecInt}) and the
aging correlation function (Sec. \ref{SecAGE}) are obtained. Sec.
\ref{SecSum} is a summary.

\section{Stochastic Model, and Definitions\label{SecModel}}

\begin{figure}
\includegraphics[%
  width=1.0\columnwidth,
  keepaspectratio]{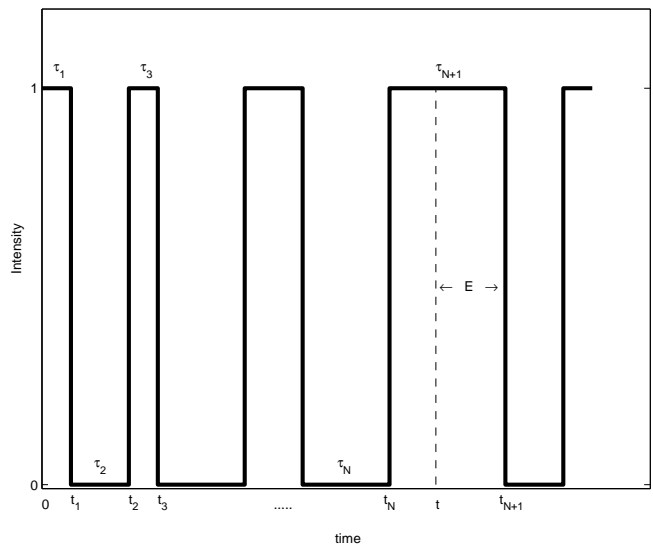}

\caption{\label{cap:figure-def}Schematic temporal evolution of the dichotomous
intensity process.}
\end{figure}

The random process considered in this manuscript, is schematically
depicted in Fig. \ref{cap:figure-def}. The intensity $I(t)$ jumps
between two states $I(t)=+1$ and $I(t)=0$. At start of the measurement
$t=0$ the NC is in state \emph{on}: $I(0)=1$. The sojourn time $\tau_{i}$
is an \emph{off} time if $i$ is even, it is an \emph{on} time if
$i$ is odd (see Fig. \ref{cap:figure-def}). The times $\tau_{i}$
for odd {[}even{]} $i$, are drawn at random from the probability
density function (PDF) $\psi_{+}(t)$, $[\psi_{-}(t)]$, respectively.
These sojourn times are mutually independent, identically distributed
random variables. Times $t_{i}$ are cumulative times from the process
starting point at time zero till the end of the \emph{i}'th transition.
Time \emph{t} on Fig. \ref{cap:figure-def} is the time of observation.

We denote the Laplace transform of $\psi_{\pm}(t)$ using \begin{equation}
\hat{\psi}_{\pm}(s)=\int_{0}^{\infty}\psi_{\pm}(t)e^{-st}\textrm{d}t.\end{equation}
 We will classify behaviors of observables of interest using the small
$s$ expansion of $\psi_{\pm}\left(s\right)$. We will consider:\\
 (i) \textbf{Case 1} PDFs with finite mean \emph{on} and \emph{off}
times, whose Laplace transform in the limit $s\rightarrow0$ satisfies:
\begin{equation}
\hat{\psi}_{\pm}(s)=1-s\tau_{\pm}+\cdots.\label{eq0m1}\end{equation}
 Here $\tau_{+}$ $(\tau_{-})$ is the average \emph{on} (\emph{off})
time. For example exponentially distributed \emph{on} and \emph{off}
times, \begin{equation}
\hat{\psi}_{\pm}(s)=\frac{1}{1+s\tau_{\pm}},\label{EqExp}\end{equation}
 belong to this class of PDFs.\\
 (ii) \textbf{Case 2} PDFs with infinite mean \emph{on} and \emph{off}
times, namely PDFs with power law behavior satisfying \begin{equation}
\psi_{\pm}\propto t^{-1-\alpha_{\pm}}\,\,\,\,\,\,\,\,\,\,\,\,\alpha_{-}<\alpha_{+}<1,\end{equation}
 in the limit of long times. The small $s$ behavior of these family
of functions satisfies \begin{equation}
\hat{\psi}_{\pm}(s)=1-A_{\pm}s^{\alpha_{\pm}}+\cdots\label{eqm0}\end{equation}
 where $A_{\pm}$ are parameters which have units of time$^{\alpha}$.
We will also consider cases where \emph{on} times have finite mean
$(\alpha_{+}=1)$ while the \emph{off} mean time diverges $(\alpha_{-}<1)$
since this situation describes behavior of uncapped QD \cite{Oijen}.
\\
 (iii) \textbf{Case 3} PDFs with infinite mean with $\alpha_{+}=\alpha_{-}=\alpha$\begin{equation}
\hat{\psi}_{\pm}(s)=1-A_{\pm}s^{\alpha}+\cdots\label{eqm0a}\end{equation}
 Note that Brokmann et al. \cite{Brokmann} report that for CdSe dots,
$\alpha_{+}=0.58\pm0.17$, and $\alpha_{-}=0.48\pm0.15$, hence within
error of measurement, $\alpha\simeq0.5$.\\
 (iv) In Sec. \ref{secDef} we will briefly consider the behavior
of the correlation function for defected $\psi_{-}(t)$.

\subsection{Physical Meaning of $\psi_{\pm}(t)$\label{sub:Physical-Meaning}}

As mentioned in Introduction, and following Ref. \cite{EfrosRosen97},
we assume that a charged (neutral) uncapped NC is in state \emph{off}
(\emph{on}), respectively. Physically, for charged NC, Auger non-radiative
decay time of a laser excited electron-hole pair, is much faster than
the radiative time of the electron-hole pair \cite{Shimuzu}. Hence
a charged NC is in \emph{off} state. The physical mechanism responsible
for the \emph{power law} blinking (i.e. charging) behavior of NCs
is still unclear. Models based on trapping of charge carriers in the
vicinity of the NC, and fluctuating barrier concepts were suggested
in \cite{Kuno,Ken,Oijen}. Here we will emphasize an alternative simple
picture based on diffusion concepts. Before further experiments are
performed, it is impossible to say if the simple picture we consider
here works better or worse than other approaches. 

We note that the simplest diffusion controlled chemical reaction $A+B\rightleftharpoons AB$,
where $A$ is fixed in space, can be used to explain some of the observed
behavior on the uncapped NCs. As mentioned the latter exhibit exponential
distribution of \emph{on} times and power law distribution of \emph{off}
times. The \emph{on} times follow standard exponential kinetics corresponding
to an ionization of a neutral NC (denoted as \emph{AB}). A model for
this exponential behavior was given already in \cite{EfrosRosen97}.
Once the NC is ionized ($A+B$ state) we assume the ejected charge
carrier exhibits a random walk on the surface of the NC or in the
bulk. This part of the problem is similar to Onsager's classical problem
of an ion pair escaping neutralization (see e.g., \cite{HongNoolandi78,SanoTachiya79}).
The survival probability in the \emph{off} state for time \emph{t},
$S_{-}(t)$ is related to the \emph{off} time distribution via $S_{-}(t)=1-\int_{0}^{t}\psi_{-}(\tau)d\tau$,
or \begin{equation}
\psi_{-}(t)=-\frac{dS_{-}(t)}{dt}.\end{equation}
 It is well known that in three dimensions survival probability decays
like $t^{-1/2}$, the exponent 1/2 is close to the exponent measured
in the experiments. In infinite domain the decay is not to zero, but
the 1/2 appears in many situations, for finite and infinite systems,
in completely and partially diffusion controlled recombination, in
different dimensions, and can govern the leading behavior of the survival
probability for orders of magnitude in time \cite{SanoTachiya79,NadlerStein91,NadlerStein96}.
In this picture the exponent $1/2$ does not depend on temperature,
similar to what is observed in experiment. We note that it is possible
that instead of the charge carrier executing the random walk, diffusing
lattice defects which serve as a trap for charge carrier are responsible
for the blinking behavior of the NCs. 

One of the possible physical pictures explaining blinking of capped
NCs can be based on diffusion process, using a variation of a three
state model of \cite{Oijen}. As mentioned in Introduction, for this
case power law distribution of \emph{on} and \emph{off} times are
observed. In particular, neutral capped NC will correspond to state
\emph{on} (as for uncapped NCs). However, capped NC can remain \emph{on}
even in the ionized state. We assume that the ionized capped NC can
be found in two states: (i) the charge remaining in the NC can be
found in center of NC (possibly a de-localized state), (ii) charge
remaining in the NC can be trapped in vicinity of capping. For case
(i) the NC will be in state \emph{off}, for case (ii) the NC will
be in state \emph{on}. The main idea is that the rate of Auger nonradiative
recombination \cite{EfrosRosen97} of consecutively formed electron-hole
pairs will drop for case (ii) but not for case (i). We note that capping
may increase effective radius of the NC, or provide trapping sites
for the hole (e.g., recent studies by Lifshitz et al. \cite{Lifshitz04}
demonstrate that coating of NCs creates trapping sites in the interface).
Thus the \emph{off} times occur when the NC is ionized and the hole
is close to the center, these \emph{off} times are slaved to the diffusion
of the electron. While \emph{on} times occur for both a neutral NC
and for charged NC with the charge in vicinity of capping, the latter
\emph{on} times are slaved to the diffusion of the electron. In the
case of power law \emph{off} time statistics this model predicts same
power law exponent for the \emph{on} times, because both of them are
governed by the return time of the ejected electron. 

The main point we would like to emphasize is that several simple mechanisms
might be responsible for the power law statistics, and hence aging
correlation functions in single molecule experiments may turn out
to be wide spread. Beyond single molecule spectroscopy we note that
certain single ion channels \cite{NadlerStein91,GoychukHanggi01,GoychukHanggi02},
deterministic diffusion in chaotic systems \cite{ZK}, the sign of
magnetization of spin systems at criticality \cite{GL}, all exhibit
intermittency behavior, and the correlation function we obtain here
might be useful also in other fields. Hence we don't restrict our
attention to the exponent 1/2, as there are indications for other
values of $\alpha$ between 0 and 1, and the analysis hardly changes.

\subsection{Definition of Correlation Functions}

Since the process under investigation is non-ergodic, and since measurements
are made on a single molecule level, care must be taken in the definition
of averages. From a single trajectory (ST) of $I(t)$, recorded in
a time interval $(0,T)$, we may construct the time average correlation
function \begin{equation}
C_{ST}(T,t')=\frac{\int_{0}^{T-t'}I(t+t')I(t)\textrm{d}t}{T-t'}.\label{eqTA}\end{equation}
 On the other hand we may generate many intensity trajectories one
at a time, to obtain $C(t,t')$ and $g^{(2)}(t,t')$. We call $C(t,t')$
single particle averaged correlation function. For non-ergodic processes
$C_{ST}(T,t')\ne C(t,t')$ even in the limit of large $t$ and $T$.
Moreover for non-ergodic processes, even in the limit of $T\rightarrow\infty$,
$C_{ST}(T,t')$ is a random function which varies from one sample
of $I(t)$ to another. 

We stress that the single particle averaged correlation function is
not the correlation function obtained from measurement of an ensemble
of particles. To see this consider the intensity of $N$ blinking
NCs, \begin{equation}
\tilde{I}_{N}=\sum_{j=1}^{N}I_{j}(t)\end{equation}
 and $j$ is an index of the particle number. The corresponding normalized
correlation function is \begin{equation}
G^{(2)}(t,t')=\frac{\langle\tilde{I}_{N}(t)\tilde{I}_{N}(t+t')\rangle}{\langle\tilde{I}_{N}(t)\rangle\langle\tilde{I}_{N}(t+t')\rangle}.\label{eqG}\end{equation}
 If the blinking behavior of individual NCs is independent, but statistically
identical, \begin{equation}
G^{(2)}(t,t')=\frac{N\langle I(t)I(t+t')\rangle+N(N-1)\langle I(t)\rangle\langle I(t+t')\rangle}{N^{2}\langle I(t)\rangle\langle I(t+t')\rangle}.\end{equation}
 Clearly for $N=1$ we obtain the correlation function $C(t,t')$
in Eq. (\ref{eq0DEF}), while for $N\rightarrow\infty$ the intensity
does not fluctuate at all (as well known). Hence while we consider
here average over an ensemble of trajectories, we are not reconstructing
the correlation function obtained from a measurement of a large number
of NCs. Our theory is valid only for single molecule measurements,
and the aging behavior of the normalized correlation function cannot
be obtained from macroscopic measurement.

\section{Number of Jump Events Between $0$ and $t$\label{SecNumber}}

In this Section we investigate basic statistical properties of the
\emph{on-off} process. 

The probability of $n$ transitions (either \emph{off} $\rightarrow$
\emph{on} or \emph{on} $\rightarrow$ \emph{off}) between times 0
and \emph{t} is \begin{equation}
P_{t}(n)\equiv P(0,t,n)=\langle\theta\left(t_{n}<t<t_{n+1}\right)\rangle\label{eq01}\end{equation}
 where $\theta(t_{n}<t<t_{n+1})$ is 1 if the event in the parenthesis
occurs; otherwise it is zero. Laplace transforming Eq. (\ref{eq01})
with respect to \emph{t} yields \begin{equation}
\hat{P}_{s}(n)=\langle\int_{t_{n}}^{t_{n+1}}\textrm{d}te^{-st}\rangle=\langle e^{-st_{n}}\left(\frac{1-e^{-s\tau_{n+1}}}{s}\right)\rangle.\label{eq02}\end{equation}
 A simple calculation using $t_{n}=\sum_{i=1}^{n}\tau_{i}$ yields
\begin{equation}
\hat{P}_{s}\left(n\right)=\left\{ \begin{array}{ll}
\left[\hat{\psi}_{+}\left(s\right)\hat{\psi}_{-}\left(s\right)\right]^{n/2}\frac{1-\hat{\psi}_{+}\left(s\right)}{s} & n\,\,\,\,\,\,\mbox{even}\\
\,\, & \,\,\\
\left[\hat{\psi}_{+}\left(s\right)\right]^{\frac{n+1}{2}}\left[\hat{\psi}_{-}\left(s\right)\right]^{\frac{n-1}{2}}\frac{1-\hat{\psi}_{-}\left(s\right)}{s} & n\,\,\,\,\,\,\mbox{odd}\end{array}\right.\label{eq03}\end{equation}
 To derive Eq. (\ref{eq03}) we used the initial condition that the
state of the process at $t=0$ is $+$. Eq. (\ref{eq03}) satisfies
the normalization condition $\sum_{n=0}^{\infty}\hat{P}_{s}(n)=1/s$.

\subsection{Mean number of renewals $\langle n\rangle$ }

Using Eq. (\ref{eq03}) the mean number of transitions is in Laplace
$t\rightarrow s$ space \begin{equation}
\langle\hat{n}(s)\rangle=\sum_{n=0}^{\infty}n\hat{P}_{s}(n)=\frac{\hat{\psi}_{+}(s)\left[1+\hat{\psi}_{-}(s)\right]}{s\left[1-\hat{\psi}_{-}(s)\hat{\psi}_{+}(s)\right]}.\label{eqavns}\end{equation}
 Using Eqs. (\ref{eq0m1}, \ref{eqm0}, \ref{eqm0a}), the small $s$
expansion of Eq. (\ref{eqavns}), and then inverting to time domain
we get the long time behavior \begin{equation}
\langle n(t)\rangle\sim\left\{ \begin{array}{cc}
\frac{2t}{\tau_{+}+\tau_{-}} & \mbox{case}\,\,\,\,1\\
\,\, & \,\,\\
\frac{2t^{\alpha_{-}}}{A_{-}\Gamma\left(1+\alpha_{-}\right)} & \mbox{case}\,\,\,\,2\\
\,\, & \,\,\\
\frac{2t^{\alpha}}{\left(A_{+}+A_{-}\right)\Gamma\left(1+\alpha\right)} & \mbox{case}\,\,\,\,3.\end{array}\right.\label{eqave}\end{equation}

\subsection{Asymptotes of $P_{t}(n)$}

For narrow PDFs, i.e. case 1, and for long times we obtain from Eqs.
(\ref{eq0m1}, \ref{eq03}) \begin{equation}
P_{t}(n)\simeq\left\{ \begin{array}{cc}
\frac{\langle\tau_{+}\rangle}{\langle\tau_{+}\rangle+\langle\tau_{-}\rangle}\delta\left(\frac{n}{2}-\frac{t}{\langle\tau_{+}\rangle+\langle\tau_{-}\rangle}\right) & n\,\,\,\,\,\,\,\,\mbox{even}\\
\,\, & \,\,\\
\frac{\langle\tau_{-}\rangle}{\langle\tau_{+}\rangle+\langle\tau_{-}\rangle}\delta\left(\frac{n+1}{2}-\frac{t}{\langle\tau_{+}\rangle+\langle\tau_{-}\rangle}\right) & n\,\,\,\,\,\,\,\,\mbox{odd}\end{array}\right.\label{eq04}\end{equation}
 To obtain this result we used the small $s$ expansion of Eq. (\ref{eq03})
and then a simple Laplace inversion. We neglected the fluctuations
in this treatment, the latter are expected to be Gaussian in the long
time limit. 

For broad PDFs satisfying $\alpha_{+}=\alpha_{-}=\alpha$, case $3$,
we find \begin{equation}
\hat{P}_{s}(n)\sim A_{\pm}s^{\alpha-1}e^{-k\left(A_{+}+A_{-}\right)s^{\alpha}}\end{equation}
 where $k=n/2$ and $A_{\pm}=A_{+}$ for $n$ even, while $k=(n+1)/2$
and $A_{\pm}=A_{-}$ for $n$ odd. Inverting to the time domain we
find \begin{equation}
P_{t}(n)\sim\frac{A_{\pm}}{\alpha}\frac{t}{(A_{+}k+A_{-}k)^{1/\alpha+1}}l_{\alpha}\left[\frac{t}{\left(A_{+}k+A_{-}k\right)^{1/\alpha}}\right].\end{equation}
 where $l_{\alpha}(t)$ is the one sided L\'{e}vy stable PDF whose
Laplace pair is $\exp(-s^{\alpha})$. 

For case $2$, with $\alpha_{-}<\alpha_{+}$, we get \begin{equation}
\hat{P}_{s}(n)\sim A_{-}\exp\left(-A_{-}ks^{\alpha_{-}}\right)\end{equation}
 for $k=n/2$ and $n$ even. The probability of finding an odd $n$
in this limit is zero. This is expected since the \emph{off} times
are much longer than the \emph{on} times, in statistical sense. Thus
for long times we have \begin{equation}
P_{t}(n)\sim\frac{A_{-}}{\alpha_{-}}\frac{t}{(A_{-}k)^{1/\alpha_{-}+1}}l_{\alpha_{-}}\left[\frac{t}{\left(A_{-}k\right)^{1/\alpha_{-}}}\right]\end{equation}
 and $n$ is even.

\section{Forward Recurrence Time\label{SecFor}}

The time $E=t_{N+1}-t$ is called the forward recurrence time. The
times (see Fig. \ref{cap:figure-def}) $t_{N+1}$ and $t_{N}$ are
defined in such a way that $t_{N}<t<t_{N+1}$, hence also $N$ is
a random variable. Let $f_{t}(E)$ be the probability density function
of the random variable $E$. The subscript $t$ in $f_{t}(E)$ indicates
that $t$ is a parameter, while $E$ is a random variable. Generally
the PDF of $E$ depends on how old the process is, namely on $t$.
A process is said to exhibit statistical aging if even in the limit
of $t\rightarrow\infty$, $f_{t}(E)$ depends on $t$. The PDF $f_{t}(E)$
is important for the calculation of the aging correlation function. 

We consider the joint PDF \begin{equation}
\textrm{f}_{t}(E,N)=\langle\delta\left(E-t_{N+1}+t\right)\theta\left(t_{N}<t<t_{N+1}\right)\rangle.\label{eqFT}\end{equation}
 Later we will sum over $N$ to obtain $f_{t}(E)$. We consider the
double Laplace transform $\mathcal{L}$ of Eq. (\ref{eqFT}) with
$t\rightarrow s$ and $E\rightarrow u$ \[
\mathcal{L}_{t,E}\left\{ \textrm{f}_{t}(E,N)\right\} =\]
 \[
=\langle\int_{t_{N}}^{t_{N+1}}\textrm{d}t\int_{0}^{\infty}\textrm{d}Ee^{-st-uE}\delta\left(E-t_{N+1}+t\right)\rangle\]

\[
=\langle e^{-ut_{N+1}}\int_{t_{N}}^{t_{N+1}}e^{-(s-u)t}\textrm{d}t\rangle\]
\begin{equation}
=\langle e^{-ut_{N+1}}\frac{e^{-(s-u)t_{N+1}}-e^{-(s-u)t_{N}}}{u-s}\rangle.\end{equation}
 For even $N$ we use the averages \[
\langle e^{-st_{N+1}}\rangle=\left[\hat{\psi}_{+}\left(s\right)\right]^{\frac{N}{2}+1}\left[\hat{\psi}_{-}\left(s\right)\right]^{\frac{N}{2}},\]
 \[
\langle e^{-st_{N}}\rangle=\left[\hat{\psi}_{+}\left(s\right)\hat{\psi}_{-}\left(s\right)\right]^{\frac{N}{2}},\]
 \[
\langle e^{-u(t_{N+1}-t_{N})}\rangle=\hat{\psi}_{+}\left(u\right),\]
 and find \begin{equation}
\hat{\textrm{f}}_{s}(u,N)=\frac{\left[\hat{\psi}_{+}\left(s\right)\psi_{-}\left(s\right)\right]^{N/2}\left[\hat{\psi}_{+}(s)-\hat{\psi}_{+}(u)\right]}{u-s}.\label{eq19}\end{equation}
 In similar way we obtain for $N$ odd \begin{equation}
\hat{\textrm{f}}_{s}(u,N)=\frac{\left[\hat{\psi}_{+}\left(s\right)\psi_{-}\left(s\right)\right]^{\frac{N-1}{2}}\left[\hat{\psi}_{-}(s)-\hat{\psi}_{-}(u)\right]\hat{\psi}_{+}(s)}{u-s}.\label{eq20}\end{equation}
 Note that $\hat{\textrm{f}}_{s}(u,N)$ is the double Laplace transform
of $\hat{\textrm{f}}_{t}(E,N)$, while $\hat{\psi}_{\pm}(s)$ and
$\hat{\psi}_{\pm}(u)$ are single Laplace transforms. Summing $\hat{\textrm{f}}_{s}(u,N)$
over $N$ we obtain the double Laplace transform of $f_{t}(E)$\begin{equation}
\hat{f}_{s}(u)=\frac{\left[\hat{\psi}_{+}\left(s\right)-\hat{\psi}_{+}\left(u\right)\right]+\hat{\psi}_{+}(s)\left[\hat{\psi}_{-}(s)-\hat{\psi}_{-}\left(u\right)\right]}{\left(u-s\right)\left[1-\hat{\psi}_{+}(s)\hat{\psi}_{-}(s)\right]}.\label{eqGL}\end{equation}
 For $\hat{\psi}_{+}(s)=\hat{\psi}_{-}(s)$ Eq. (\ref{eqGL}) reduces
to Eq. $6.2$ in \cite{GL}.

\subsection{Limiting cases for $f_{t}(E)$ }

We now analyze the long time $t\rightarrow\infty$ behavior of $f_{t}(E)$.
In this case we expect that an equilibrium PDF for $f_{t}(E)$ will
emerge. This equilibrium is related to stationarity, ergodicity, and
aging as we will show. 

We consider narrow distributions, i.e. case $1$ first. Taking the
limit $s\rightarrow0$ of Eq. (\ref{eqGL}), corresponding to $t\rightarrow\infty$
and find \begin{equation}
\hat{f}_{s}(u)\sim\frac{1-\hat{\psi}_{+}\left(u\right)}{su\left(\tau_{+}+\tau_{-}\right)}+\frac{1-\hat{\psi}_{-}\left(u\right)}{su\left(\tau_{+}+\tau_{-}\right)}.\label{eq12}\end{equation}
 The Laplace $s\rightarrow t$ and $u\rightarrow E$ inversion of
this equation is immediate \begin{equation}
f_{t}(E)=\frac{\int_{E}^{\infty}\psi_{+}(t')\textrm{d}t'}{\tau_{+}+\tau_{-}}+\frac{\int_{E}^{\infty}\psi_{-}(t')\textrm{d}t'}{\tau_{+}+\tau_{-}},\label{eq13}\end{equation}
 a behavior which is valid in the limit of long time $t$ (and independent
of it). Note that the first (second) term on the right hand side of
the equation, corresponds to trajectories with even (odd) number of
steps. One can show that in the limit of long times probability of
finding the process in state $\pm$ is \begin{equation}
\lim_{t\rightarrow\infty}P_{\pm}(t)=\frac{\tau_{\pm}}{\tau_{+}+\tau_{-}},\label{eq1314}\end{equation}
 as might be expected. In the special case of $\psi_{-}=\psi_{+}$
we obtain a well known equation \cite{Feller} which has several applications
in theory of random walks e.g. \cite{Haus}. The important point to
notice is that in the limit of large time $t$, and when average times
$\tau_{\pm}$ are finite, an equilibrium is obtained which does not
depend on $t$. 

We now consider broad distributions, with diverging averaged \emph{on}
and \emph{off} times, case $2$. In the limit of small $s$ and small
$u$, with their ratio finite \begin{equation}
\hat{f}_{s}(u)\sim\frac{u^{\alpha_{-}}-s^{\alpha_{-}}}{(u-s)s^{\alpha_{-}}}.\label{eq14}\end{equation}
 The investigation of this equation yields the long time $t$ behavior
\begin{equation}
f_{t}(E)\sim d_{t}(E),\label{eq15}\end{equation}
 where $d_{t}(E)$ is Dynkin's function \begin{equation}
d_{t}(E)=\frac{\sin(\pi\alpha_{-})}{\pi}\frac{t^{\alpha_{-}}}{E^{\alpha_{-}}(t+E)}.\label{eq15Dyn}\end{equation}
 From Eqs. (\ref{eq14},\ref{eq15}) we learn that unlike case 1,
the PDF of $E$ depends on time $t$ even in the long time limit.
Eq. (\ref{eq15Dyn}) was obtained by Dynkin \cite{Feller,Dynkin}
as a limit theorem for renewal processes with a single waiting time
PDF. Here we showed that for a two state process the details on $\alpha_{+}$
and $A_{+}$ are not important in the long time limit. This is expected,
the \emph{off} (i.e. minus) times are much longer than the \emph{on}
(i.e. plus) times in statistical sense, and hence our results in the
long time limit are not sensitive to the details of $\psi_{+}(t)$.
In the same spirit it can be shown that in the limit of long time
$t$, and with probability one, the process is found in state minus. 

Finally for case $3$ where $\alpha_{+}=\alpha_{-}<1$, we find that
Eqs. (\ref{eq14},\ref{eq15}) are still valid. However now probability
of finding the process in state $\pm$ is given by \begin{equation}
P_{\pm}\equiv\lim_{t\rightarrow\infty}P_{\pm}(t)=\frac{A_{\pm}}{A_{+}+A_{-}}.\label{eq:Pplusminus}\end{equation}

\subsection{Joint PDFs for Forward Recurrence time}

It will turn out important to define the joint PDFs of time $E$ provided
that process is in state plus or state minus at time $t$. We denote
these PDFs with $f_{t}(E,\pm)$ and the corresponding double Laplace
transform $\hat{f}_{s}(u,\pm)$. Since the start of process is state
$+$ at time $t=0$, we get using Eq. (\ref{eq19}) \[
\hat{f}_{s}(u,+)=\sum_{N=0,N\mbox{ even}}^{\infty}\hat{\textrm{f}}_{s}(u,N)=\]
\begin{equation}
\frac{\hat{\psi}_{+}(s)-\hat{\psi}_{+}(u)}{(u-s)\left[1-\hat{\psi}_{+}(s)\hat{\psi}_{-}(s)\right]}.\label{eqJoint}\end{equation}
 and using Eq. (\ref{eq20}) \[
\hat{f}_{s}(u,-)=\sum_{N=0,N\mbox{ odd}}^{\infty}\hat{\textrm{f}}_{s}(u,N)=\]
\begin{equation}
\frac{\hat{\psi}_{+}(s)\left[\hat{\psi}_{-}(s)-\hat{\psi}_{-}(u)\right]}{(u-s)\left[1-\hat{\psi}_{+}(s)\hat{\psi}_{-}(s)\right]}.\end{equation}
 Note that \begin{equation}
\hat{f}_{s}(u)=\hat{f}_{s}(u,-)+\hat{f}_{s}(u,+).\end{equation}
 The probability of finding the particle in state $+$ when $t\rightarrow\infty$
is \begin{equation}
\lim_{t\rightarrow\infty}P_{+}=\lim_{t\rightarrow\infty}\int_{0}^{\infty}f_{t}(E,+)\textrm{d}E,\label{eq16}\end{equation}
 provided that the limit exists. For example for case 1 it is easy
obtain from Eq. (\ref{eq16}) the result in Eq. (\ref{eq1314}). 

The limiting PDFs $f_{t}(E,\pm)$ are obtained in double Laplace space
by considering the small $s$ (and small \emph{u} for cases 2 and
3) limit. They are \begin{equation}
\hat{f}_{s}(u,\pm)\sim\left\{ \begin{array}{cc}
\frac{1-\hat{\psi}_{\pm}(u)}{us(\tau_{+}+\tau_{-})} & \mbox{case}\,\,1\\
\,\, & \,\,\\
\frac{A_{\pm}}{A_{+}+A_{-}}\frac{u^{\alpha}-s^{\alpha}}{(u-s)s^{\alpha}} & \mbox{case}\,\,3.\end{array}\right.\label{eqcc13}\end{equation}
 For case $2$ we find in this limit $\hat{f}_{s}(u,+)=0$, i.e. probability
of finding the particle in state $+$ is zero, and \begin{equation}
\hat{f}_{s}(u,-)=\frac{u^{\alpha}-s^{\alpha}}{(u-s)s^{\alpha}}.\end{equation}
 The double inverse Laplace transform of this equation is given in
Eq. (\ref{eq15Dyn}).

\section{Number of Renewals Between Two Times\label{SecRen}}

We now calculate $P(t,t+t',n)$ the probability of number of renewals
$n$ between time $t$ and time $t+t'$. Obviously the process is
generally not stationary and the information on $P(0,t',n)$, obtained
in Sec. \ref{SecNumber}, is not sufficient for the determination
of $P(t,t+t',n)$. We now classify the trajectories according to the
state of the process (i.e., $+$ or $-$) at times $t$ and $t+t'$.
It will turn out that the intensity trajectories, when the process
is in state $+$ at time $t$ and state $+$ at time $t+t'$, are
those which are important for the calculation of the correlation function. 

The probability of not making a jump in time interval $t,t+t'$, when
the process is in state $+$ at time $t$ and state $+$ at time $t+t'$
is \begin{equation}
P_{++}(t,t+t',0)=\int_{t'}^{\infty}f_{t}(E,+)\textrm{d}E.\label{eqpp0}\end{equation}
 The probability of finding $n>0$ transition events in time interval
$t,t+t'$, when state of process at time $t$ is $+$ and state of
process is $t+t'$ is also $+$ \[
\mathcal{L}_{t'}P_{++}(t,t+t',n)=\]
\begin{equation}
\hat{f}_{t}(u,+)\hat{\psi}_{-}(u)\left[\hat{\psi}_{+}(u)\hat{\psi}_{-}(u)\right]^{\frac{n}{2}-1}\frac{1-\hat{\psi}_{+}(u)}{u},\label{eqpp}\end{equation}
 where $n>0$ is even, and $u$ is the Laplace conjugate of $t'$.
Note that Eq. (\ref{eqpp}) depends on $t$ through $\hat{f}_{t}(u,+)$.
The other combinations, e.g. $P_{+-}(t,t+t',n)$, are given in Appendix
\ref{Appendix0}, as well as $P(t,t+t',n)$.

\section{Mean Intensity of \emph{on-off} process\label{SecInt}}

The averaged intensity $\langle I(t)\rangle$ for the process switching
between 1 and 0 and starting at 1 is now considered. In Laplace $t\rightarrow s$
space it is easy to show that \begin{equation}
\left\langle \hat{I}(s)\right\rangle =\frac{1-\hat{\psi}_{+}(s)}{s}\cdot\frac{1}{1-\hat{\psi}_{+}(s)\hat{\psi}_{-}(s)}.\label{EqMeanIntensity}\end{equation}
 One method to obtain this equation is to note that $\langle I(t)\rangle=\mbox{Prob}[I(t)=1]$,
hence \begin{equation}
\left\langle \hat{I}(s)\right\rangle =\int_{0}^{\infty}f_{s}(E,+)\textrm{d}E,\end{equation}
 and therefore the $u\rightarrow0$ limit of Eq. (\ref{eqJoint})
yields Eq. (\ref{EqMeanIntensity}). 

The Laplace $s\rightarrow t$ inversion of Eq. (\ref{EqMeanIntensity})
yields the mean intensity $\langle I(t)\rangle$. Using small $s$
expansions of Eq. (\ref{EqMeanIntensity}), we find in the limit of
long times \begin{equation}
\langle I(t)\rangle\sim\left\{ \begin{array}{cc}
\frac{\tau_{+}}{\tau_{+}+\tau_{-}} & \mbox{case}\,\,\,\,1\\
\,\, & \,\,\\
\frac{A_{+}t^{\alpha_{-}-\alpha_{+}}}{A_{-}\Gamma\left(1+\alpha_{-}-\alpha_{+}\right)} & \mbox{case}\,\,\,\,2\\
\,\, & \,\,\\
\frac{A_{+}}{A_{+}+A_{-}} & \mbox{case}\,\,\,\,3.\end{array}\right.\label{eqIave}\end{equation}

If the \emph{on} times are exponential, as in Eq. (\ref{EqExp}) then
\begin{equation}
\left\langle \hat{I}(s)\right\rangle =\frac{\tau_{+}}{1+s\tau_{+}-\psi_{-}(s)}.\label{eq:Iexpon}\end{equation}
 This case corresponds to the behavior of the uncapped NCs. The expression
in Eq. (\ref{eq:Iexpon}), and more generally, the case $\alpha_{-}<\alpha_{+}=1$
leads for long time \emph{t} to \begin{equation}
\left\langle I(t)\right\rangle \sim\frac{\tau_{+}t^{\alpha_{-}-1}}{A_{-}\Gamma(\alpha_{-})}.\label{eqiii}\end{equation}
 For exponential \emph{on} and \emph{off} time distributions Eq. (\ref{EqExp}),
we obtain the exact solution \begin{equation}
\left\langle I(t)\right\rangle =\frac{\tau_{-}\exp\left[-t\left(\frac{1}{\tau_{-}}+\frac{1}{\tau_{+}}\right)\right]+\tau_{+}}{\tau_{-}+\tau_{+}}.\label{eqExp}\end{equation}

\textbf{Remark} For the case $\alpha_{+}<\alpha_{-}<1$, corresponding
to a situation where \emph{on} times are in statistical sense much
longer then \emph{off} times, $\langle I(t)\rangle\sim1$.

\section{Aging Correlation Function of \emph{on}-\emph{off} process\label{SecAGE}}

We are now able to calculate the correlation function $C(t,t')=\langle I(t)I(t+t')\rangle$.
We consider the process $I(t)$ as jumping between state \emph{on}
with $I(t)=+1$ and state \emph{off} $I(t)=0$. The symmetric case
where $I(t)$ jumps between the states $I(t)=-1$ or $I(t)=-1$, is
discussed in Appendix \ref{Appendix1}. We assume that sojourn times
in state \emph{on} (\emph{off}) are described by $\psi_{+}(t)$ ($\psi_{-}(t)$),
respectively. Contributions to the correlation function arise only
from trajectories with $I(t)=1$ and $I(t+t')=1$, meaning that only
the $++$ trajectories, in Eq. (\ref{eqpp}) contribute to the correlation
function. Summing Eq. (\ref{eqpp}) over even $n>0$, and using Eq.
(\ref{eqpp0}) for $n=0$, we find \[
\hat{C}(t,u)=\frac{\hat{f}_{t}(u=0,+)-\hat{f}_{t}(u,+)}{u}\]
\begin{equation}
+\hat{f}_{t}(u,+)\frac{\hat{\psi}_{-}(u)\left[1-\hat{\psi}_{+}(u)\right]}{u\left[1-\hat{\psi}_{-}(u)\hat{\psi}_{+}(u)\right]},\label{eq:Corr}\end{equation}
 where $u$ is the Laplace conjugate of $t'$. We see that the correlation
function generally depends on time $t$.

\subsection{Case 1}

For case $1$ with finite $\tau_{+}$ and $\tau_{-}$, and in the
limit of long times $t$, we find \[
\lim_{t\rightarrow\infty}\hat{C}(t,u)=\]
\begin{equation}
\frac{1}{u}\frac{\tau_{+}}{\tau_{+}+\tau_{-}}\left\{ 1-\frac{\left[1-\hat{\psi}_{+}(u)\right]\left[1-\hat{\psi}_{-}(u)\right]}{\tau_{+}u\left[1-\hat{\psi}_{-}(u)\hat{\psi}_{+}(u)\right]}\right\} \label{eqVerO}\end{equation}
 This result was obtained by Verberk and Orrit \cite{Verberk} and
it is seen that the correlation function depends asymptotically only
on $t'$(since \emph{u} is Laplace pair of $t'$). Namely, when average
\emph{on} and \emph{off} times are finite the system does not exhibit
aging. If both $\psi_{+}(t)$ and $\psi_{-}(t)$ are exponential then
the \emph{exact} result is\[
\begin{array}{ccc}
C(t,t') & = & \frac{\tau_{-}\exp\left[-t\left(\frac{1}{\tau_{-}}+\frac{1}{\tau_{+}}\right)\right]+\tau_{+}}{\tau_{-}+\tau_{+}}\\
\\ & \times & \frac{\tau_{-}\exp\left[-t'\left(\frac{1}{\tau_{-}}+\frac{1}{\tau_{+}}\right)\right]+\tau_{+}}{\tau_{-}+\tau_{+}}\end{array}\]
 and $C(t,t')$ becomes independent of \emph{t} exponentially fast
as \emph{t} grows.

\subsection{Case 2}

We consider case $2$, however limit our discussion to the case $\alpha_{+}=1$
and $\alpha_{-}<1$. As mentioned this case corresponds to uncapped
NCs where \emph{on} times are exponentially distributed, while \emph{off}
times are described by power law statistics. Using the exact solution
Eq. (\ref{eq:Corr}) we find asymptotically, when both \emph{both
t} and $t'$ are large: \begin{equation}
C(t,t')\sim\left(\frac{\tau_{+}}{A_{-}}\right)^{2}\frac{\left(tt'\right)^{\alpha_{-}-1}}{\Gamma^{2}\left(\alpha_{-}\right)}.\label{eqG1}\end{equation}
 Unlike case $1$ the correlation function approaches zero when $t\rightarrow\infty$,
since when $t$ is large we expect to find the process in state \emph{off}.
Using Eq. (\ref{eqiii}), the asymptotic behavior of the normalized
correlation function Eq. (\ref{eq:g2def}) is \begin{equation}
g^{(2)}(t,t')\sim\left(1+\frac{t}{t'}\right)^{1-\alpha_{-}}.\label{eq:case2g2asymp}\end{equation}
 We see that the correlation functions Eqs. (\ref{eqG1}, \ref{eq:case2g2asymp})
exhibit aging, since they depend on the age of the process $t$. 

Considering the asymptotic behavior of $C(t,t')$ for large \emph{t},
but small $t'$, yields in the limit of $s\ll u,\, s\ll u\hat{\psi}(u)$
\[
\hat{C}(t,u)\approx\]
\begin{equation}
\frac{1}{u}\frac{\tau_{+}}{A_{-}\Gamma(\alpha_{-})t^{1-\alpha_{-}}}\left\{ 1-\frac{\left[1-\hat{\psi}_{+}(u)\right]\left[1-\hat{\psi}_{-}(u)\right]}{\tau_{+}u\left[1-\hat{\psi}_{-}(u)\hat{\psi}_{+}(u)\right]}\right\} .\label{eqG11}\end{equation}
 This equation is similar to Eq. (\ref{eqVerO}), especially if we
notice that the {}``effective mean'' time of state \emph{off} until
total time \emph{t} scales as $A_{-}t^{1-\alpha_{-}}$. Despite assumptions
of $s\ll u,\, s\ll u\hat{\psi}(u)$ in the derivation of Eq. (\ref{eqG11}),
it also reproduces the result Eq. (\ref{eqG1}) and hence is applicable
for any \emph{u} (and thus $t'$) as long as \emph{t} is large enough. 

For the special case, where \emph{on} times are exponentially distributed,
the correlation function \emph{C} is a product of two identical expressions
\emph{for all} \emph{t} and $t'$: 

\begin{equation}
\hat{C}(s,u)=\frac{\tau_{+}}{1+s\tau_{+}-\psi_{-}(s)}\cdot\frac{\tau_{+}}{1+u\tau_{+}-\psi_{-}(u)},\label{eqtwoid}\end{equation}
 where \emph{s} (\emph{u}) is the Laplace conjugate of \emph{t} (\emph{t'})
respectively. Comparing to Eq. (\ref{eq:Iexpon}) we obtain \begin{equation}
C(t,t')=\langle I(t)\rangle\langle I(t')\rangle,\label{eqcprod}\end{equation}
 and for the normalized correlation function \begin{equation}
g^{(2)}(t,t')=\frac{\left\langle I(t')\right\rangle }{\left\langle I(t+t')\right\rangle }.\label{eqg2ratio}\end{equation}
 Eqs. (\ref{eqg2ratio}, \ref{eqcprod}) are important since they
show that measurement of mean intensity $\langle I(t)\rangle$ yields
the correlation functions, for this case. While our derivation of
Eqs. (\ref{eqg2ratio}, \ref{eqcprod}) is based on the assumption
of exponential on times, it is valid more generally for any $\psi_{+}(t)$
with finite moments, in the asymptotic limit of large $t$ and $t'$.
To see this note that Eqs. (\ref{eqG1}, \ref{eqiii}) yield $C(t,t')\sim\langle I(t)\rangle\langle I(t')$. 

\begin{figure}
\includegraphics[%
  width=1.0\columnwidth,
  keepaspectratio]{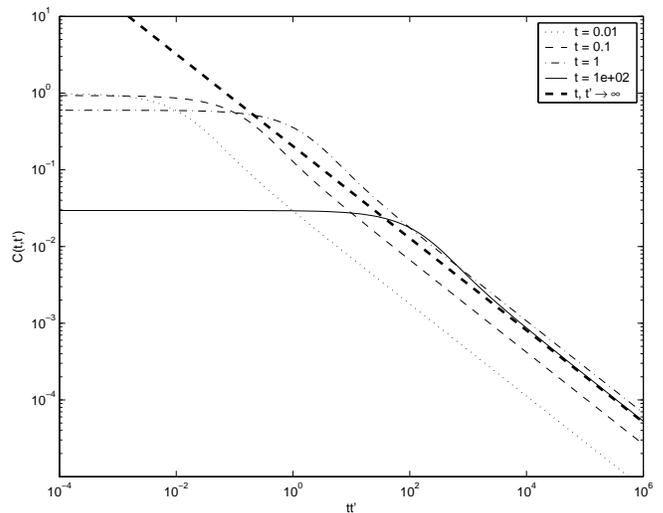}

\caption{\label{fig3} Exact $C(t,t')$ for Case $2$: exponential \emph{on}
times and power law \emph{off} times with $\alpha_{-}=0.4$. We use
$\hat{\psi}_{+}(s)=1/(1+s)$ and $\hat{\psi}_{-}(s)=1/(1+s^{0.4})$
and numerically obtain the correlation function. For each curve in
the figure we fix the time $t$. The process starts in the state $on$.
Thick dashed straight line shows the asymptotic behavior Eq. (\ref{eqG1}).
For short times ($t'<1$ for our example) we observe the behavior
$C(t,t')\sim C(t,0)=\langle I(t)\rangle$, the correlation function
is flat.}
\end{figure}

In Fig. \ref{fig3} we compare the asymptotic result (\ref{eqG1})
with exact numerical double Laplace inversion of the correlation function.
We use exponential PDF of \emph{on} times $\psi_{+}(s)=1/(1+s)$,
and power law distributed \emph{off} times: $\hat{\psi}_{-}(s)=\hat{\psi}_{-}(s)=1/(1+s^{0.4})$
corresponding to $\alpha_{-}=0.4$. Convergence to asymptotic behavior
is observed. 

\textbf{Remark} For fixed $t$ the correlation function in Eq. (\ref{eqG1})
exhibits a $(t')^{\alpha_{-}-1}$ decay. A $(t')^{\alpha_{-}-1}$
decay of an intensity correlation function was reported in experiments
of Orrit's group \cite{Oijen} for uncapped NCs (for that case $\alpha_{-}=0.65\pm0.2$).
However, the measured correlation function is a time averaged correlation
function Eq. (\ref{eqTA}) obtained from a single trajectory. In that
case the correlation function is independent of $t$, and hence no
comparison between theory and experiment can be made yet.

\subsection{Case 3}

We now consider case $3$, and for long \emph{t} and $t'$ we find
\begin{equation}
C(t,t')=P_{+}-P_{+}P_{-}\frac{\sin\pi\alpha}{\pi}B\left(\frac{1}{1+t/t'};1-\alpha,\alpha\right)\label{eq:case3asympcorr}\end{equation}
 with $P_{\pm}$ given by Eq. (\ref{eq:Pplusminus}) and where\[
B(z;a,b)=\int_{0}^{z}x^{a-1}(1-x)^{b-1}dx\]
 is the incomplete beta function. The behavior in this limit does
not depend on the detailed shape of the PDFs of the \emph{on} and
\emph{off} times, besides the parameters $A_{+}/A_{-}$ and $\alpha$
(see also Appendix \ref{Appendix2}). We note that both terms of Eq.
(\ref{eq:Corr}) contribute to Eq. (\ref{eq:case3asympcorr}). The
appearance of the incomplete beta function in Eq. (\ref{eq:case3asympcorr})
is related to the concept of persistence. The probability of not switching
from state \emph{on} to state \emph{off} in a time interval $(t,t+t')$,
assuming the process is in state \emph{on} at time $t$, is called
the persistence probability. In the scaling limit this probability
is found using Eqs. (\ref{eq15Dyn}, \ref{eqcc13}): \[
\frac{\sin(\pi\alpha)}{\pi}\int_{t'}^{\infty}\frac{t^{\alpha}}{E^{\alpha}(t+E)}\textrm{d}E\]
\begin{equation}
=1-\frac{\sin\pi\alpha}{\pi}B\left(\frac{1}{1+t/t'};1-\alpha,\alpha\right)\label{eqPers}\end{equation}
 The persistence implies that long time intervals in which the process
does not jump between states \emph{on} and \emph{off}, control the
asymptotic behavior of the correlation function. The factor $P_{+}$,
which is controlled by the amplitude ratio $A_{+}/A_{-}$, determines
the expected short and long time $t'$ behaviors of the correlation
function, namely $C(\infty,0)=\lim_{t\rightarrow\infty}\langle I(t)I(t+0)\rangle=P_{+}$
and $C(\infty,\infty)=\lim_{t\rightarrow\infty}\langle I(t)I(t+\infty)\rangle=(P_{+})^{2}$.
With slightly more details the two limiting behaviors are: \begin{equation}
C(t,t')\sim\left\{ \begin{array}{cc}
P_{+} & \,\,\,\frac{t'}{t}\ll1\\
\,\, & \,\,\\
(P_{+})^{2}+P_{+}\frac{\sin(\pi\alpha)}{\pi\alpha}\left(\frac{t'}{t}\right)^{-\alpha} & \,\,\,\frac{t'}{t}\gg1.\end{array}\right.\label{eq:case3asympasymp}\end{equation}
 Using Eq. (\ref{eqIave}) the normalized intensity correlation function
is $g^{(2)}(t,t')\sim C(t,t')/(P_{+})^{2}$. 

\begin{figure}
\includegraphics[%
  width=1.0\columnwidth,
  keepaspectratio]{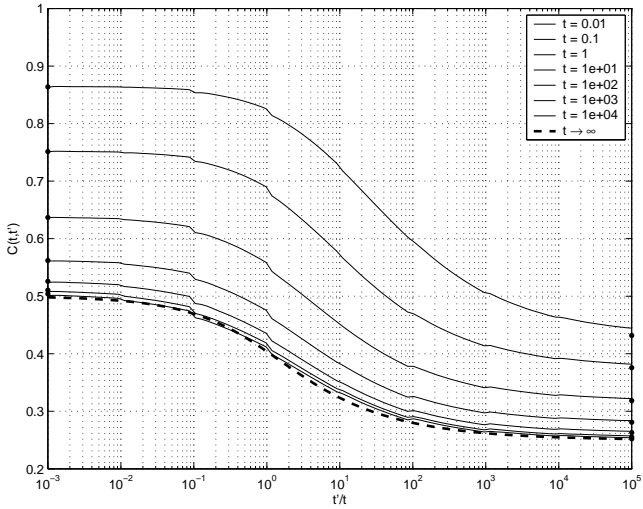}

\caption{\label{cap:alpha0.4exact}Exact $C(t,t')$ for case $3$, when both
\emph{on} and \emph{off} times are power law distributed with $\alpha=0.4$.
We use $\hat{\psi}_{\pm}(s)=1/(1+s^{0.4})$ for different times \emph{t}
increasing from the topmost to the lowermost curves. The dots on the
left and on the right show $C(t,0)=\left\langle I(t)\right\rangle $
and $C(t,\infty)=\left\langle I(t)\right\rangle /2$ respectively.
The process starts in the state \emph{on}.}
\end{figure}

In Fig. \ref{cap:alpha0.4exact} we compare the asymptotic result
(\ref{eq:case3asympcorr}) with exact numerical double Laplace inversion
of the correlation function for PDFs $\hat{\psi}_{+}(s)=\hat{\psi}_{-}(s)=1/(1+s^{0.4})$.
Convergence to Eq. (\ref{eq:case3asympcorr}) is seen. 

\textbf{Remark} For small $t'/t$ we get flat correlation functions.
Flat correlation functions were observed by Dahan's group \cite{Dahan}
for capped NCs. However, the measured correlation function is a single
trajectory correlation function Eq. (\ref{eqTA}), and hence no comparison
between theory and experiment can be made yet.

\subsection{Defected \emph{off} time distribution\label{secDef}}

As mentioned in the Introduction, an \emph{off} state of uncapped
NC corresponds to an ionized NC. Assume that the transition from state
\emph{on} to state \emph{off} occurs when a charge carrier is ejected
into the vicinity of the NC, and then starts to move diffusively in
the bulk. If the diffusion process takes place in three dimensions,
there is a finite probability that the charge carrier will not return
to the NC. In that case the NC remains in state \emph{off} forever. 

Such a situation can be modeled based on defected distribution of
\emph{off} times. In this case we have a non-normalized PDF of \emph{off}
times \begin{equation}
\int_{0}^{\infty}\psi_{-}(t)dt=Z<1,\label{eq:psioffdefect}\end{equation}
 the small $s$ expansion of the Laplace transform of $\psi_{-}(t)$
is $\hat{\psi}(s)\approx Z-A_{-}s^{\alpha_{-}}$. $Z$ is the probability
of charge carrier to return; this probability was the subject of extensive
investigation in the context of first passage time problems \cite{Redner}. 

For large \emph{t} the mean intensity is \begin{equation}
\left\langle I(t)\right\rangle \sim\left\{ \begin{array}{cc}
\frac{A_{+}t^{-\alpha_{+}}}{(1-Z)\Gamma(1-\alpha_{+})}, & \alpha_{\pm}<1\\
\\\frac{A_{+}A_{-}t^{-(1+\alpha_{-})}}{(1-Z)^{2}|\Gamma(-\alpha_{-})|}, & \,\,\,\alpha_{-}<\alpha_{+}=1.\end{array}\right.\label{eqDefI}\end{equation}
 Note that here $\alpha_{-}$ can be smaller, larger or equal to $\alpha_{+}$
when $\alpha_{+}<1$. 

Using Eq. (\ref{eq:Corr}) we obtain asymptotically, for both \emph{t}
and $t'$ large,\begin{equation}
C(t,t')\sim\left\{ \begin{array}{cc}
\frac{A_{+}}{(1-Z)\Gamma(1-\alpha_{+})}(t+t')^{-\alpha_{+}}, & \alpha_{\pm}<1\\
\,\, & \,\,\\
\left(\frac{A_{-}A_{+}}{(1-Z)^{2}\Gamma(-\alpha_{-})}\right)^{2}(tt')^{-(1+\alpha_{-})}, & \alpha_{-}<\alpha_{+}=1.\end{array}\right.\label{eq:Corrdefect}\end{equation}
 Using Eq. (\ref{eqDefI}) we can relate the intensity correlation
function with the mean intensity, \begin{equation}
C(t,t')\sim\left\{ \begin{array}{cc}
\langle I(t+t')\rangle, & \alpha_{\pm}<1\\
\,\, & \,\,\\
\langle I(t)\rangle\langle I(t')\rangle, & \,\,\,\alpha_{-}<\alpha_{+}=1.\end{array}\right.\label{eq:CorrdefectInt}\end{equation}

This result shows that for $\alpha_{+}<1$ $\left\langle I(t)I(t+t')\right\rangle \simeq\left\langle I(t+t')\right\rangle $,
independent of the value of $I(t)$. Asymptotic validity of this relation
can be explained by noticing that the non-zero contributions to $\left\langle I(t)I(t+t')\right\rangle $
come only when both $I(t)$ and $I(t+t')$ are equal to 1. However,
at long time \emph{t}, after jumping \emph{off} there is a negligible
probability of being \emph{on} again for a cumulative time duration
comparable with the total time (note that $\left\langle I(t)\right\rangle $
scales as the probability of making no transition \emph{off}, i.e.,
of persistence). Hence, the nonzero contributions to $\left\langle I(t+t')\right\rangle $
are mainly those staying \emph{on} from time \emph{t}, so that almost
certainly, if $I(t+t')=1$ then also $I(t)=1$ and $\left\langle I(t)I(t+t')\right\rangle \simeq\left\langle I(t+t')\right\rangle $. 

The above argument does not hold in the case of $\alpha_{+}=1$, because
now the \emph{on} times are very short and there is no possibility
of staying \emph{on} persistently for long times. Accordingly, the
decay of $\left\langle I(t)\right\rangle $ and $C(t,t')$ is much
faster here. The leading contributions to $\left\langle I(t)\right\rangle $
and $C(t,t')$ in the case of $\alpha_{+}<1$ disappear as $\alpha_{+}\nearrow1$,
due to the pole of the $\Gamma$-function.

\section{Summary\label{SecSum}}

We demonstrated the dependence of the two-time correlation function
$C(t,t')=\left\langle I(t)I(t+t')\right\rangle $ on the times $t$
and $t'$. This is in a full contrast to the well-known convergence
of the correlation function to the stationary limit which is independent
of \emph{t}. Such a convergence is found when the average \emph{on}
and \emph{off} times are finite (as shown above for exponential \emph{on}
and \emph{off} distributions). When these times diverge non-stationary
behavior is found. The non-vanishing \emph{t}-dependence of the correlation
function $C(t,t')$ is known as aging. 

We obtain different modes of aging, yielding dependence of $C(t,t')$
on the ratio $t/t'$, product $tt'$ and the sum $t+t'$.\\
 (i) For PDF of \emph{on} times having finite mean and power law distributed
\emph{off} times with infinite mean, the correlation function asymptotically
splits into a product of two identical functions, one of \emph{t}
and the other of $t'$ (see Eq. (\ref{eqtwoid})), leading to $tt'$
dependence Eq. (\ref{eqG1}). This case corresponds to the behavior
of the uncapped NCs. \\
 (ii) When both \emph{on} and \emph{off} times are described by broad
distributions, with identical exponents $\alpha_{+}=\alpha_{-}$,
the correlation function depends on the ratio $t/t'$, Eq. (\ref{eq:case3asympcorr}).
This case corresponds to the capped NCs (within the error of measurement).\\
 (iii) For defected \emph{off} times and $\alpha_{+}<1$, we find
that the correlation function depends on $t+t'$, Eq. (\ref{eq:Corrdefect}).\\
 (iv) Finally, for stochastic processes with finite \emph{on} and
\emph{off} times, we recover known behavior, where the correlation
function in the scaling limit depends only on $t'$, Eq. (\ref{eqVerO}).\\

In different regimes, the correlation function exhibits either a strong
sensitivity on the details of the stochastic process (i.e., on $\psi_{\pm}(t)$),
or certain universal features which are now discussed. We also found
relations between the correlation function and mean intensity, for
several cases.\\
 (i) For PDF of \emph{on} times having finite mean and power law distributed
\emph{off} times with infinite mean, the correlation function is related
to the mean intensity according to $C(t,t')\sim\langle I(t)\rangle\langle I(t')\rangle$
Eq. (\ref{eqcprod}). For short times $t'$ the correlation function
depends on the details of $\psi_{\pm}(t)$, Eq. (\ref{eqG11}). \\
 (ii) When both \emph{on} and \emph{off} times are described by broad
distributions, with identical exponents $\alpha_{+}=\alpha_{-}$,
the persistence probability governs the aging correlation function
Eq. (\ref{eq:case3asympcorr}). This is a universal behavior in the
sense that all $\psi_{\pm}(t)$ belonging to this family, yield identical
behavior for the correlation function $C(t,t')$ in the limit of $t\rightarrow\infty$.
\\
 (iii) For defected \emph{off} times and $\alpha_{+}<1$, we find
that the correlation function $C(t,t')=\langle I(t+t')\rangle$.\\
 (iv) For the standard case, where both the mean \emph{on} and mean
\emph{off} times are finite, the correlation function depends on the
details of $\psi_{\pm}(t)$.

Simple physical explanations for the aging behavior were briefly discussed.
Models, discussed previously in the literature, based on diffusion
processes, or fluctuating barrier models, or trap models, may all
lead to aging behavior of the correlation function. Thus aging behaviors
in single molecule spectroscopy may have other applications besides
single nano-crystal spectroscopy. The dependence of the correlation
function on control parameters like temperature and laser intensity
can be used to distinguish between the microscopic scenarios proposed
here and in the literature. 

\begin{acknowledgments}
Acknowledgment is made to the National Science Foundation for support
of this research. EB acknowledges fruitful discussions with M. Bawendi,
K. Kuno, and M. Orrit, as well as J. P. Bouchaud for his mini-course
on anomalous processes. 
\end{acknowledgments}
\appendix

\section{Number of Transitions Between $t$ and $t+t'$\label{Appendix0}}

The probability of finding $n>0$ transition events in time interval
$t,t+t'$, when the process is in state $+$ at time $t$, and state
$-$ at time $t+t'$ \[
\mathcal{L}_{t'}P_{+-}(t,t+t',n)=\]
\begin{equation}
\hat{f}_{t}(u,+)\left[\hat{\psi}_{+}(u)\hat{\psi}_{-}(u)\right]^{\frac{n-1}{2}}\frac{1-\hat{\psi}_{-}(u)}{u},\label{eqpm}\end{equation}
 $n$ is odd. The probability of finding $n=0$ transition events
in time interval $t,t+t'$, when the process is in state $-$ at time
$t$, and state $-$ at time $t+t'$\begin{equation}
P_{--}(t,t+t',0)=\int_{t'}^{\infty}f_{t}(E,-)\textrm{d}E,\end{equation}
 with obvious notations, and for even $n$ \[
\mathcal{L}_{t'}P_{--}(t,t+t',n)=\]
\begin{equation}
\hat{f}_{t}(u,-)\hat{\psi}_{+}(u)\left[\hat{\psi}_{+}(u)\hat{\psi}_{-}(u)\right]^{\frac{n}{2}-1}\frac{1-\hat{\psi}_{-}(u)}{u}.\label{eqmm}\end{equation}
 Finally for odd $n$ \[
\mathcal{L}_{t'}P_{-+}(t,t+t',n)=\]
\begin{equation}
\hat{f}_{t}(u,-)\left[\hat{\psi}_{+}(u)\hat{\psi}_{-}(u)\right]^{\frac{n-1}{2}}\frac{1-\hat{\psi}_{+}(u)}{u}.\label{eqmp}\end{equation}
 The indexes $ij=++$, $ij=--$, $ij=-+$, and $ij=+-$ correspond
to trajectories which start in state $i$ at time $t$ and end with
state $j$ at time $t+t'$. Obviously for $P(t,t+t',n)$ we have for
even $n>0$\begin{equation}
\mathcal{L}_{t'}P(t,t+t',n)=\mathcal{L}_{t'}P_{++}(t,t+t',n)+\mathcal{L}_{t'}P_{--}(t,t+t',n),\end{equation}
 while for odd $n$\begin{equation}
\mathcal{L}_{t'}P(t,t+t',n)=\mathcal{L}_{t'}P_{+-}(t,t+t',n)+\mathcal{L}_{t'}P_{-+}(t,t+t',n).\end{equation}

\section{jumps between 1 and -1\label{Appendix1}}

We now consider a correlation function which slightly differs than
the one considered in the main text. We assume that $I(t)=+1$ or
$I(t)=-1$, the $+$ $[-]$ times are described by $\psi_{+}(t)$
$[\psi_{-}(t)]$ respectively. For this case the correlation function
is related to $P(t,t+t',n)$ according to \begin{equation}
C(t,t')=\sum_{n=0}^{\infty}(-1)^{n}P(t,t+t',n).\end{equation}
 In Laplace $t'\rightarrow u$ space, and using convolution theorem
of Laplace transform, \[
\mathcal{L}_{t'}C(t,t')=\hat{C}(t,u)=\]
 \[
\frac{\hat{f}_{t}(u=0,+)-\hat{f}_{t}(u,+)}{u}+\frac{\hat{f}_{t}(u=0,-)-\hat{f}_{t}(u,-)}{u}+\]
 \[
\sum_{n>0,\,\,\mbox{even}}\mathcal{L}_{t'}P_{++}(t,t+t',n)+\sum_{n>0,\,\,\mbox{even}}\mathcal{L}_{t'}P_{--}(t,t+t',n)\]
\begin{equation}
-\sum_{n>0,\,\,\mbox{odd}}\mathcal{L}_{t'}P_{-+}(t,t+t',n)-\sum_{n\,\,\mbox{odd}}\mathcal{L}_{t'}P_{+-}(t,t+t',n).\label{eqfor}\end{equation}
 The first two terms on the right hand side of equation (\ref{eqfor})
correspond to trajectories with no transitions in time interval $t,t+t'$.
The $\hat{f}_{t}(u=0,\pm)$ are the probabilities $P_{\pm}$ of finding
the process in state $\pm$ at time $t$. Using Eqs. (\ref{eqpp},
\ref{eqpm}, \ref{eqmm}, \ref{eqmp}) we get \[
\hat{C}(t,u)=\frac{1-\sum_{i=\pm}\hat{f}_{t}(u,i)}{u}+\]
\begin{equation}
\frac{1}{u\left[1-\hat{\psi}_{-}(u)\hat{\psi}_{+}(u)\right]}\sum_{i=\pm}\hat{f}_{t}(u,i)\left[\hat{\psi}_{i^{*}}(u)(2-\hat{\psi}_{i}(u))-1\right],\end{equation}
 where $i^{*}=+$ if $i=-$ and $i^{*}=-$ if $i=+$. If \emph{on}
and \emph{off} times are identically distributed, $\psi_{+}(t)=\psi_{-}(t)$
we obtain the result given by Godr\'{e}che and Luck \cite{GL}.

~

~

\section{Asymptotics of $C(t,t')$ for case 3 ($\alpha_{+}=\alpha_{-}\equiv\alpha<1$)
\label{Appendix2}}

If, in analogy to the cases 1 and 2, we wish to explore the behavior
of the correlation function in the limit of large \emph{t}, but for
any time difference $t'$, it is easy to obtain the following asymptotic
result for small \emph{s} from Eq. (\ref{eq:Corr}): \begin{equation}
\hat{C}(s,u)\approx\frac{1}{u}P_{+}\left\{ \frac{1}{s}-\frac{1-\hat{\psi}_{-}(u)}{(s-u)\left[1-\hat{\psi}_{-}(u)\hat{\psi}_{+}(u)\right]}\right\} .\label{eq:Csuapp}\end{equation}
 However, consistent with the demand $t'\ll t$ we have $u\gg s$
and so have to remove \emph{s} in the second term in the brackets
of Eq. (\ref{eq:Csuapp}). It would be wrong to try and attempt inverse
Laplace inversion of this term with respect to \emph{s}. Thus, we
obtain\[
C(t,t'\ll t)\approx P_{+},\]
 in agreement with Eq. (\ref{eq:case3asympasymp}). We see that the
correlation function is virtually constant for any $t'$ (even small)
and for any \emph{t} large enough, as long as $t'\ll t$. This, of
course, could be expected based on the fact that \emph{asymptotic}
expression Eq. (\ref{eq:case3asympcorr}) gives the exact value of
$P_{+}$ for $t'=0$ (see also Fig. \ref{cap:alpha0.4exact}). 

Eq. (\ref{eq:Csuapp}) can also be used to check the asymptotics when
$t'$ also becomes large. Using small \emph{u} expansions for $\hat{\psi}_{\pm}(u)$
yields\[
\hat{C}(s,u)\approx P_{+}\left\{ \frac{1}{su}-\frac{P_{-}}{u(s-u)}\right\} .\]
 The double inverse Laplace transform of $1/u(s-u)$ is either 0 or
1 (for positive $t,t'$), depending on whether we assume $u>s$ or
$u<s$ (i.e., do we first perform inversion with respect to \emph{u}
or to \emph{s}). The choice $u>s$ is appropriate when $t'\ll t$
and vice versa, hence we recover asymptotic limits shown in Eq. (\ref{eq:case3asympasymp}),
up to the leading order. 

To conclude, we have demonstrated that in case 3, in the long \emph{t}
limit the correlation function does not depend on the particular form
of $\psi_{\pm}$ but only on their asymptotics, for any $t'$, as
given by Eq. (\ref{eq:case3asympcorr}). This is in contrast to the
long \emph{t} limiting behavior of $C(t,t')$ in cases 1 and 2, where
$C(t,t')$ does depend on the particular form of $\psi_{\pm}$ for
short $t'$.

\end{document}